# Identification of memory kernels in thermo-viscoelasticity

M. Sellier

*Schott AG – Hattenbergstrasse 10, 55122 Mainz, Germany*
*URL: www.schott.com                    e-mail: mathieu.sellier@itwm.fraunhofer.de*

ABSTRACT: This paper discusses the possibility of identifying the shear and structure relaxation kernels in glassy materials by means of a single, simple and non-intrusive experiment. The material should be thermorheologically simple and the kernels expressed in the form of Prony series. The experiment considered consists in measuring the thickness variations over time of a flat sample cooled symmetrically from both side from a temperature above the glass transition temperature down to room temperature. The comparison of experimental observations with theoretically calculated responses allows the identification of the coefficients of the Prony series for the shear and structure relaxation kernels using a least-square type method. This paper illustrates the success of the method with 'artificially' created experimental observations and with up to two exponential terms in the Prony series for the shear and structure relaxation kernels.

Key words: parameter identification, viscoelasticity, stress relaxation, structure relaxation, thermorheologically simple, glass

## 1 INTRODUCTION

Many numerical techniques are nowadays well established to simulate the thermo-mechanical behaviour of viscoelastic material in complex geometries and under non-trivial loading conditions. The potential ability of these techniques to reproduce the reality and therefore to provide useful predictive results depends strongly on the quality of the estimate of the material properties.

Materials such as glass undergo stress relaxation upon loading at temperatures around the glass transition. They are also prone to structure relaxation upon cooling, i.e. depending on the cooling path the state of the structure characterized by the fictive temperature will differ. Both relaxation phenomena are well described by means of Boltzmann memory integrals [1] and the challenge is to identify the time-dependent mechanical moduli functions and the time-dependent structure relaxation function (memory kernels). This task is often tackled in practice by performing separate experiments for the time-dependent mechanical moduli functions on one hand and for the structure relaxation function on the other hand. For example, torsion test is used to identify the shear modulus function at a given temperature and thermorheological simplicity is invoked to extrapolate this function for other temperatures. A review of the possible experimental procedures is given in [2]. Likewise, the structure relaxation function is found by imposing a temperature jump to the material and monitoring a material property variation [1].

The experiment considered here consists in measuring the thickness variations over time of a flat sample cooled symmetrically from both side from a temperature above the glass transition temperature down to room temperature. Comparison of experimental observations with theoretically predicted thickness variation allows the identification of the kernels coefficients. Thus, the next section describes the governing equations dictating the sample thickness variations and briefly discusses how these are solved. The Levenberg-Marquardt method used to identify the kernels parameters is then described and preliminary results with 'artificially' created experimental observations are presented.

## 2 EVALUATION OF THE SAMPLE THICKNESS VARIATION

## 2.1 Constitutive equations

As a standard starting point of the thermo-mechanical analysis (see [1,3,4]), the glassy material is pre-supposed to obey the following constitutive equations:

$$s_{ij}(x,t) = 2\int_0^t G(\xi(x,t)-\xi(x,t'))\frac{\partial e_{ij}(x,t')}{\partial t'}dt',$$
$$\sigma(x,t) = 3K(\varepsilon(x,t)-\varepsilon_{th}(x,t)),$$
(1)

where $s_{ij}$, $\sigma$ and $e_{ij}$, $\varepsilon$ are the deviatoric and volumetric parts of the stress and strain tensor respectively, $K$ the constant bulk modulus and the shear modulus $G$ is function of the elapsed reduced time $\xi(x,t)-\xi(x,t')$. If moreover the influence of the temperature on the relaxation behaviour can be represented by a classical Arrhenius model, the reduced time is expressed as, [5]:

$$\xi(x,t) = \int_0^t \frac{\tau_{ref}}{\tau(T,T_f)}dt' = \int_0^t e^{\frac{\Delta H}{R}\left(\frac{1}{T_0}-\frac{1}{2T}-\frac{1}{2T_f}\right)}dt',$$
(2)

where $\tau_{ref}$ and $\tau(T,T_f)$ are the relaxation times at the reference temperature $T_0$ and the temperature $T$ respectively. $T_f$ is the fictive temperature which characterizes the state of the structure, $\Delta H$ the activation energy and $R$ the ideal gas constant. The shear modulus and fictive temperature can be expressed in the form of Prony series as follows,

$$G(\xi) = G_\infty + \sum_{i=1}^n G_i e^{-\xi/\lambda_i},\ G_i = v_i(G_0-G_\infty),\ \sum_{i=1}^n v_i = 1,$$
$$T_f(x,t) = \sum_{i=1}^m \omega_i T_{f_i}(x,t),\ \sum_{i=1}^m \omega_i = 1.$$
(3)

In eqs. (3), $G_0$ and $G_\infty$ are the initial and final shear moduli respectively, $v_i$ and $\omega_i$ are weights to be determined and $\lambda_i$ are the unknown constants associated with a discrete relaxation spectrum in shear. $T_{fi}$ are the partial fictive temperatures introduced in [6] to ease the computation of the fictive temperature. These must satisfy the following ODE:

$$\frac{dT_{f_i}}{dt} = -\frac{T_{f_i}-T}{\mu_i}\frac{d\xi}{dt},$$
(4)

where $\mu_i$ are unknown constants associated with a discrete structural relaxation spectrum. Finally, the thermal strain in (1) is given by:

$$\varepsilon_{th} = \alpha_g(T-T_0)+(\alpha_l-\alpha_g)(T_f-T_0),$$
(5)

where $\alpha_g$, $\alpha_l$ = coefficient of thermal expansion of the solid and liquid glass respectively.

## 2.2 Reduction to a one-dimensional problem

If the glassy material sample is flat and has an initial thickness $2b_0$ much smaller than its lateral extent, the thermo-mechanical problem can be treated as one-dimensional with quantities only depending on the thickness coordinate (z on figure 1) and time t.

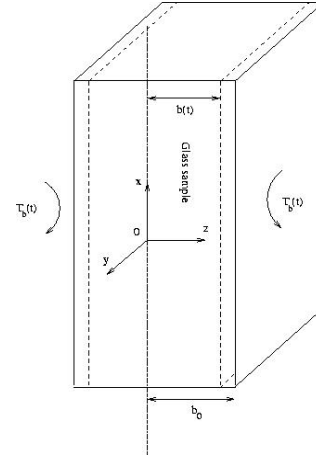

Fig 1: Sketch of the geometry and notations.

The sample has initially a uniform temperature $T_0$ and is subject to symmetric Dirichlet boundary conditions on both sides. Accordingly, if radiative heat transfer is neglected, the temperature field must satisfy

$$\begin{cases}\rho c\frac{\partial T}{\partial t} = k\left(\frac{\partial^2 T}{\partial z^2}\right),\ T(z,t=0)=T_0\\ \left.\frac{\partial T}{\partial z}\right|_{z=0} = 0,\ T(z=b_0,t) = \max(T_0-qt, 293K)\end{cases}$$
(6)

where standard notations are used and $q$ is the constant cooling rate.

The one-dimensional assumption and symmetry of the cooling considerably reduce the complexity of the mechanical problem. Because the sample is traction free, the in-plane stress must satisfy

$$\int_0^{b_0}\sigma_{xx}(z,t) = 0.$$
(7)

Moreover, for reasons of symmetry well discussed in [3,7], stresses and strains are related as follows:

$$\sigma_{xx}(z,t) = 2\int_0^t G(\xi-\xi')\frac{\partial}{\partial t'}[\varepsilon_{xx}(t')-\varepsilon_{zz}(z,t')]dt',$$
(8)

$$\sigma_{zz} = 0 = \int_0^t -\frac{4}{3}G(\xi-\xi')\frac{\partial}{\partial t'}[\varepsilon_{xx}(t')-\varepsilon_{zz}(z,t')] +$$
$$K\frac{\partial}{\partial t'}[2\varepsilon_{xx}(t')+\varepsilon_{zz}(z,t')-3\varepsilon_{th}(z,t')]dt'.$$
(9)

Finally, the thickness of the sample is obtained using

$$b(t) = b_0 + \int_0^{b_0}\varepsilon_{zz}(z,t)dz.$$
(10)

## 2.3 Solution technique

For conciseness reasons, only a brief overview of the numerical technique adopted to solve this system of coupled equations is reported here. The heat equation (6) is first solved using a standard Finite Difference technique and the Crank-Nicholson time integration scheme. The fictive temperature is then obtained by integrating eq. (4) using the stable scheme proposed in [6] and substituting in eq. (3). Equation (2) is then integrated with the Trapezoidal Rule to obtain the reduce time. Equations (8) and (9) are discretized in a way similar to that of Taylor et al., [8]. The benefit of this scheme is that it does not require the entire strain history to compute the memory integrals saving the computational resources. $\Delta\varepsilon_{zz}$ (normal strain variation for the current time step) is eliminated from the discrete analogue of eq. (8) using eq. (9) so that $\Delta\varepsilon_{xx}$ can be evaluated subject to eq. (7).

Results were validated by comparing the residual stresses with results of the fully three-dimensional simulation performed with the commercial Finite Element code Ansys. Differences of no more than 5% were achieved in a fraction of the computational time required by Ansys.

## 3 KERNEL PARAMETERS IDENTIFICATION

The definition of the inverse (identification) problem is the following: 'identify the shear and structure relaxation parameters ($v_i$, $\lambda_i$, $\omega_i$, $\mu_i$) in eqs. (3) and (4) so that the calculated response (sample thickness variation) matches in a least square sense experimental observations', i.e.:

$$\min_{p \in P_{adm}} E(p) = \frac{1}{2}\sum_{i=1}^{l} r_i^2(p) = \frac{1}{2}\sum_{i=1}^{l}\left(b(t_i) - b^{obs}(t_i)\right)^2 \quad (11)$$

where $b^{obs}$=observed thickness variation and $p$=the vector parameter. The constraints on the parameters are that $\lambda_i$ and $\mu_i$ should be strictly positive ($\forall i$), and the sum of the weights $v_i$ and $\omega_i$ should be equal to one. The popular Levenberg-Marquardt method which combines the benefits of the Newton and steepest descent methods is used to minimize eq. (11). It determines iteratively the necessary correction $dp^k$ to the vector parameter $p$ so that:

$$\left[\left(J^{(k)}\right)^T J^{(k)} + \beta^{(k)} I\right] dp^k = -\left(J^{(k)}\right)^T r^{(k)}, \quad (12)$$

$J^{(k)}$ is the Jacobian matrix of $E$ calculated using Finite Differences and $\beta^{(k)}$ is a positive constant chosen equal to $10^{-3}$ initially. If the current iteration successfully decreases $E$, $\beta^{(k)}$ is divided by 10. It is multiplied by 10 otherwise. The vector parameter is updated as follows,

$$p^{(k+1)} = p^{(k)} + dp^{(k)}. \quad (13)$$

In order to transform the constrained least-square problem into an unconstrained one, a variable transformation is performed which depends on the number of terms considered in the Prony series:

- <u>One term in the Prony series</u>: the only unknowns are $\lambda_1$ and $\mu_1$ since $v_1$ and $\omega_1$ are necessarily equal to one. The choice of vector parameter $p=(p_1,p_2)^T$ with $(p_1^2,p_2^2)^T=(\lambda_1,\mu_1)^T$ naturally enforce the constraints.
- <u>Two terms in the Prony series</u>: 6 unknowns need to be identified ($\lambda_1,\lambda_2,\mu_1, \mu_2,v_1, \omega_1$). Providing $0<v_1<1$ and $0<\omega_1<1$, the remaining two weights ($v_2$ and $\omega_2$) are found using the condition that the sum of the weights should equal one. The choice of vector parameter $p=(p_1,p_2,p_3,p_4,p_5,p_6)^T$ with $(p_1^2,p_2^2,p_3^2,p_4^2,\sin^2(p_5),\sin^2(p_6))^T=(\lambda_1,\lambda_2,\mu_1, \mu_2,v_1, \omega_1)^T$ ensures that the constraints are satisfied.

## 4 PRELIMINARY RESULTS

Clearly, a larger number of terms in the Prony series will lead to a better fit with the true kernels. However, as revealed by the last section, the number of parameters grows quickly. Results are presented here for up to two terms in the Prony series. True experimental results are not yet available. Therefore, in order to assess the method, artificial tests are performed. The experimental thickness variation $b^{obs}(t_i)$ in eq. (11) is obtained numerically for particular values of $(v_i, \lambda_i, \omega_i, \mu_i)^T = (v_i^{obs}, \lambda_i^{obs}, \omega_i^{obs}, \mu_i^{obs})^T$ and for $b_0$=1 cm, $T_0$ = 873.15 K, $q$=1°/s, $K$=4.32x$10^{10}$ Pa, $G_0$=2.85x$10^{10}$ Pa, $G_\infty$=$10^{-6}$ Pa, $\alpha_l$=40x$10^{-6}$ K$^{-1}$, $\alpha_g$=10x$10^{-6}$ K$^{-1}$, $\rho$=2537 kg/m$^3$, $c$=1320 J/kgK, $k$=1 W/mK. Then, starting from an arbitrary point in the parameter space, the identification method is applied and the convergence to these values checked.

### 4.1 One term in the Prony series

For all tested cases, the convergence to ($\lambda_1^{obs},\mu_1^{obs}$) was achieved regardless of the starting point in the parameter space. As an illustrative example, figure 2 reports the convergence history (ratio of the identified relaxation times to the 'experimental' ones as a function of the number of iterations) of the Levenberg-Marquardt method for ($\lambda_1^{obs},\mu_1^{obs}$)=(0.0095,0.308) and for a starting point ($10^{-5},10^{-5}$) orders of magnitude away from the experimental values. The number of observation points ($l$ in eq. (11)) is chosen equal to 100 and 26 iterations are necessary to recover ($\lambda_1^{obs},\mu_1^{obs}$) and reduce $E$ below $10^{-20}$.

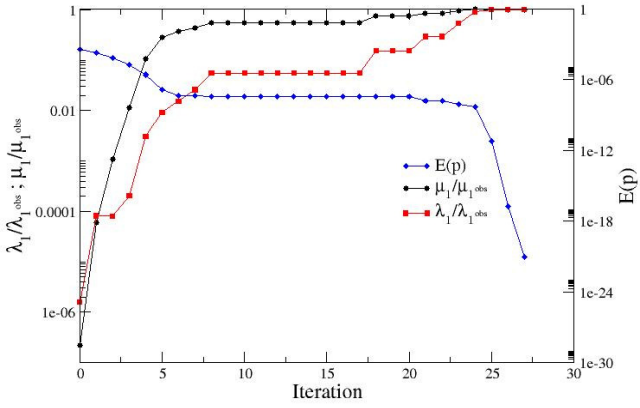

Fig 2: Convergence history of the Levenberg-Marquardt method for one term in the Prony series.

### 4.2 Two terms in the Prony series

In order to assess the method with two terms in the Prony series, values of ($\lambda_1^{obs}, \lambda_2^{obs}, \mu_1^{obs}, \mu_2^{obs}, \nu_1^{obs}, \omega_1^{obs}$) are chosen so that the resulting kernels fit the ones of a commonly used glass. Accordingly, these parameters are chosen equal to (0.0528, 0.00271, 0.411, 0.00271, 0.29, 0.77).

Moreover, artificial noise is added to the observed thickness variation, so that:

$$b^{obs,mag}(t) = (2mag*rand(t) + (1-mag))b^{obs}(t), rand(t) \in [0,1] \quad (14)$$

Results are summarized on figure 3 for three values of *mag*: 0, 1 μm and 2 μm and 500 observation points. Since the overall thickness variation is around 80 μm, these three cases correspond to an uncertainty on the observed sample thickness of 0%, 2.5% and 5% respectively. Regardless of the magnitude of the noise, the glass thickness variations shown on the upper figure is very closely recovered. The lower two figures demonstrate that in the absence of noise (*mag*=0), both memory kernels are perfectly identified and the 'experimental' parameters are precisely recovered by the algorithm.

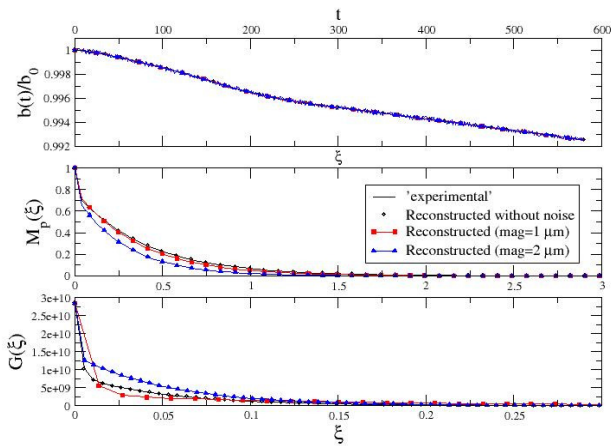

Fig 3: Thickness variation, structure relaxation kernel and shear relaxation kernel.

By definition, $M_p(\xi) = \omega_1 e^{(-\xi/\mu_1)} + (1-\omega_1)e^{(-\xi/\mu_2)}$ on the figure in the middle. Not surprisingly, as the magnitude of the noise increases, the quality of the reconstructed kernels degrades. Nevertheless, their main features such as the initial steep decrease are still adequately captured by the method.

## 5 CONCLUSIONS

This paper demonstrates, at least in theory, the possibility to identify the shear modulus and structure relaxation kernels through the measurement of the time variation of the glass sample thickness. Only comparison with real experimental data can guaranty the viability of the method but a number of potential benefits may be outlined. The method is non-intrusive, it allows the identification of the shear modulus and structure relaxation kernels through a single experiment and most importantly, the thickness variation can be measured in practice with a high degree of accuracy and with little noise. At least two points would deserve further attention. The first relates to the radiative heat transfer neglected in the present study but likely to contribute to the heat transfer at such high temperatures. The second concerns the addition of exponential terms in the Prony series. Tests with three terms in the Prony series leading to 10 unknown coefficients were successfully performed but the convergence is, at present, too sensitive to the choice of initial guess in the parameters space.